\documentclass[%
 reprint,
superscriptaddress,
 amsmath,amssymb,
 aps,
prd
]{revtex4-2}

\usepackage{xspace}

\usepackage{graphicx}
\usepackage{dcolumn}
\usepackage{bm}
\usepackage{hyperref}

\newcommand{\Kr}[1]{\ensuremath{^{#1}\mathrm{Kr}}\xspace}

\newcommand{\Tl}[1]{\ensuremath{^{#1}\mathrm{Tl}}\xspace}
\newcommand{\Xe}[1]{\ensuremath{^{#1}\mathrm{Xe}}\xspace}
\newcommand{\Cs}[1]{\ensuremath{^{#1}\mathrm{Cs}}\xspace}
\newcommand{\Th}[1]{\ensuremath{^{#1}\mathrm{Th}}\xspace}
\newcommand{\Co}[1]{\ensuremath{^{#1}\mathrm{Co}}\xspace}
\newcommand{\Bi}[1]{\ensuremath{^{#1}\mathrm{Bi}}\xspace}
\newcommand{\K}[1]{\ensuremath{^{#1}\mathrm{K}}\xspace}

\newcommand{\bb}{\ensuremath{\beta\beta}\xspace}
\newcommand{\bbnonu}{\ensuremath{0\nu\beta\beta}\xspace}
\newcommand{\bbtwonu}{\ensuremath{2\nu\beta\beta}\xspace}
\newcommand{\Ttwonu}{\ensuremath{T_{1/2}^{2\nu}}\xspace}

\begin{document}


\title{Measurement of the \Xe{136} two-neutrino double beta decay half-life via direct background subtraction in NEXT}

\collaboration{NEXT Collaboration}



\author{P.~Novella}
\affiliation{Instituto de F\'isica Corpuscular (IFIC), CSIC \& Universitat de Val\`encia, Calle Catedr\'atico Jos\'e Beltr\'an, 2, Paterna, E-46980, Spain}
\author{M.~Sorel}
\affiliation{Instituto de F\'isica Corpuscular (IFIC), CSIC \& Universitat de Val\`encia, Calle Catedr\'atico Jos\'e Beltr\'an, 2, Paterna, E-46980, Spain}
\author{A.~Us\'on}
\thanks{Corresponding author.}
\affiliation{Instituto de F\'isica Corpuscular (IFIC), CSIC \& Universitat de Val\`encia, Calle Catedr\'atico Jos\'e Beltr\'an, 2, Paterna, E-46980, Spain}
\author{C.~Adams}
\affiliation{Argonne National Laboratory, Argonne, IL 60439, USA}
\author{H.~Almaz\'an}
\affiliation{Department of Physics, Harvard University, Cambridge, MA 02138, USA}
\author{V.~\'Alvarez}
\affiliation{Instituto de Instrumentaci\'on para Imagen Molecular (I3M), Centro Mixto CSIC - Universitat Polit\`ecnica de Val\`encia, Camino de Vera s/n, Valencia, E-46022, Spain}
\author{B.~Aparicio}
\affiliation{Department of Organic Chemistry I, University of the Basque Country (UPV/EHU), Centro de Innovaci\'on en Qu\'imica Avanzada (ORFEO-CINQA), San Sebasti\'an / Donostia, E-20018, Spain}
\author{A.I.~Aranburu}
\affiliation{Department of Applied Chemistry, Universidad del Pais Vasco (UPV/EHU), Manuel de Lardizabal 3, San Sebasti\'an / Donostia, E-20018, Spain}
\author{L.~Arazi}
\affiliation{Unit of Nuclear Engineering, Faculty of Engineering Sciences, Ben-Gurion University of the Negev, P.O.B. 653, Beer-Sheva, 8410501, Israel}
\author{I.J.~Arnquist}
\affiliation{Pacific Northwest National Laboratory (PNNL), Richland, WA 99352, USA}
% ...
%


\date{\today}

\begin{abstract}
We report a measurement of the half-life of the \Xe{136} two-neutrino double beta decay performed with a novel direct background subtraction technique. The analysis relies on the data collected with the NEXT-White detector operated with \Xe{136}-enriched and \Xe{136}-depleted xenon, as well as on the topology of double-electron tracks. With a fiducial mass of only 3.5~kg of Xe, a half-life of $2.34^{+0.80}_{-0.46}\textrm{(stat)}^{+0.30}_{-0.17}\textrm{(sys)}\times10^{21}~\textrm{yr}$ is derived from the background-subtracted energy spectrum. The presented technique demonstrates the feasibility of unique background-model-independent neutrinoless double beta decay searches.
\end{abstract}


\maketitle


\section{Introduction}


After the confirmation that neutrinos are massive particles and that lepton flavor is not conserved, double beta (\bb) decay experiments \cite{Dolinski:2019nrj} stand as the main probe to explore lepton number violation and the nature of neutrino masses \cite{Gomez-Cadenas:2011oep}. \bb decay is a second order transition occurring in some even-even nuclei, for which the $\beta$ decay is highly suppressed or energetically forbidden. In this process, two bound neutrons are simultaneously transformed into two protons plus two electrons. The decay mode in which two antineutrinos are emitted (\bbtwonu) has been directly observed in nine nuclides with half-lives in the range of $\approx$10$^{19}$--10$^{21}$~yr \cite{Barabash:2020nck}. Neutrinoless \bb decay (\bbnonu) has not been detected, with the most sensitive searches probing half-lives up to 10$^{26}$~yr \cite{KamLAND-Zen:2016pfg,GERDA:2020xhi}. The \bbnonu decay violates lepton number conservation and implies the Majorana nature of neutrinos, providing also insights into their absolute mass scale. As such, the detection of this process has become one of the major goals in particle physics.


The \bbtwonu decay in \Xe{136} has been already observed in Refs.  \cite{Albert:2013gpz} and \cite{KamLAND-Zen:2019imh}, mainly following a calorimetric approach. As for the other non-geochemical measurements in Ref. \cite{Barabash:2020nck}, the half-life of this process (\Ttwonu) has been measured relying on background models derived from the screening of the detector materials and Monte Carlo (MC) simulations. The rates of \bbtwonu and background events are extracted by comparing such models to the observed data. This background-model-dependent approach is also followed in the search for \bbnonu decay, providing results that might depend on the background assumptions adopted, such as the number, type, or spatial origin of the different sources. This arises as a possible limitation for next-generation experiments, as in a background regime of $\approx$1 count/(tonne yr) new background sources of unknown origin and/or complex modeling may become relevant. The NEXT technology offers the capability to perform a direct background subtraction, regardless of the origin or number of the specific sources, by combining \Xe{136}-enriched and \Xe{136}-depleted data. Having a negligible contribution of \Xe{137} activation (feasible also in future detectors as described in Ref. \cite{NEXT:2020qup}), the current \bbtwonu analysis represents a first proof-of-principle for virtually background-model-independent \bb searches, which could be extended to the \bbnonu mode. Even in the case of non-negligible \Xe{137} activation, this technique can be extended in future detectors to measure at the same time the \Xe{136}  \bb and \Xe{137} $\beta$ decay contributions, in the absence of any other backgrounds. In addition, beyond the energy-related observables, the detailed topology of the reconstructed tracks in NEXT is uniquely exploited to enhance the \bb signal.

\section{The NEXT-White detector}


Within the roadmap of the NEXT project \cite{NEXT:2015wlq,NEXT:2020amj} to use high-pressure electroluminescent gaseous xenon time projection chambers (TPCs) for \bbnonu searches, NEXT-White \cite{NEXT:2018rgj} represents the first radiopure, large-scale demonstrator. The detector was operated at the Laboratorio Subterr\'aneo de Canfranc from 2016 to 2021. Using xenon as both the detection medium and the source of \bb decays, charged particles interacting in the active volume produce primary (S1) and secondary (S2) scintillation light, the latter by means of electroluminescence (EL) once the ionization electrons cross a high-field region close to the anode. While the detection of the S1 light determines the initial time of the interaction, the S2 signal is used to trigger the detector and to measure the energy and topological signature of the event.

\begin{figure}
  \includegraphics[width=\columnwidth]{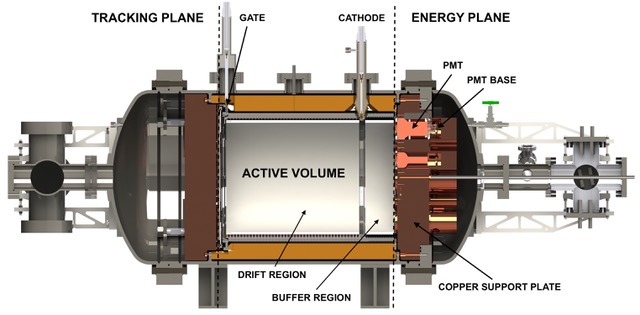}
\caption{\label{fig:new} Schematic view of the NEXT-White TPC. The drift region between the cathode and the EL gate holds $\approx$4.3 kg of xenon when operated at 10 bars. Two dedicated readout planes, for tracking and energy measurement, are placed at the extremes of the TPC.}
\end{figure}

As shown in Fig.~\ref{fig:new}, the TPC defines a cylindrical volume with an active region of 530.3~mm along the drift direction and a radius of 208~mm. When operating at 10~bars, it holds $\approx$4.3 kg of xenon. A cathode grid and a transparent anode are located at the opposite ends of the TPC. A grid (gate) defining the EL region is placed at a 6~mm distance from the anode plate. An array of 12 Hamamatsu R11410-10 3-in. photomultiplier tubes (PMTs) is located 13~cm behind the cathode. A second array of 1792 SensL series-C 1 mm$^2$ silicon photomultipliers (SiPMs) is placed 2~mm behind the anode plate. All surfaces facing the active volume are coated with tetraphenyl butadiene in order to shift the vacuum ultraviolet (VUV) light to the visible spectrum. In addition to an internal shielding made of 60--120~mm thick ultra-pure copper, two lead structures surround the pressure vessel. A radon abatement system flushes air into the space enclosed by the two lead castles, providing a virtually airborne-Rn-free environment \cite{Novella:2018ewv,Novella:2019cne}.

\begin{figure}
\includegraphics[width=\columnwidth]{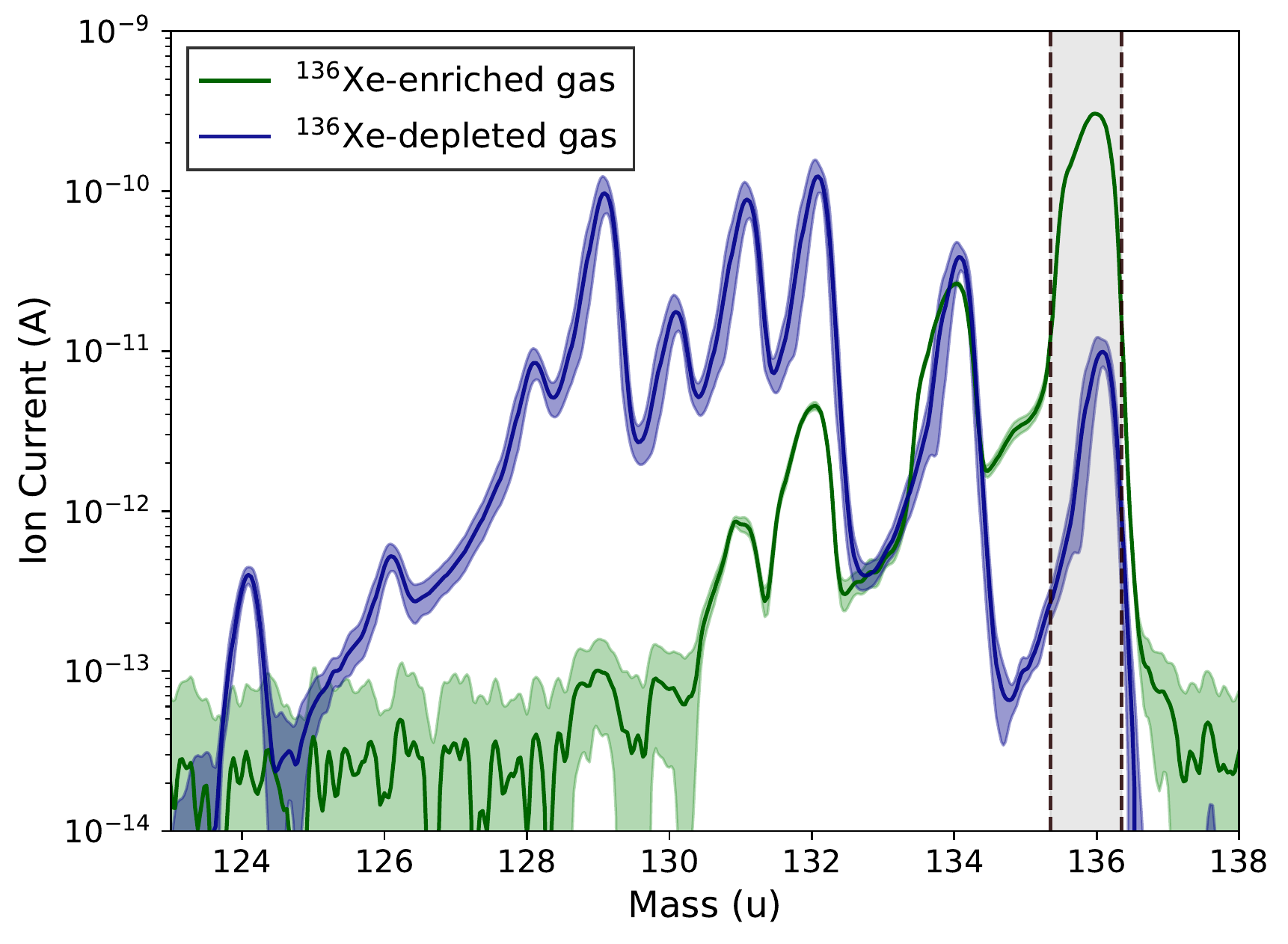}
\caption{\label{fig:isocomp} Isotopic composition of the \Xe{136}-enriched (green line) and  \Xe{136}-depleted (blue line) xenon gas. The shadowed regions around the lines correspond to the uncertainty inferred from different RGA scans. The nine stable xenon isotopes are identified. The vertical dashed lines show the 1 mass unit-wide integration region considered to derive the fractions of \Xe{136}.}
\end{figure}

The detector was operated with both xenon enriched in \Xe{136} and xenon depleted in this isotope. The isotopic compositions of the \Xe{136}-enriched and the \Xe{136}-depleted gas were measured with a residual gas analyzer (RGA), yielding \Xe{136} fractions of 90.9$\pm$0.4\% and 2.6$\pm$0.2\%, respectively. The fraction of the different xenon isotopes in both gases is presented in Fig.~\ref{fig:isocomp}. The two gases came from the same provider and as part of the same isotope separation process of natural xenon, the \Xe{136}-depleted gas constituting part of the tailings of the \Xe{136} enrichment process. Before recirculation and purification within the detector, all gas bottles were certified for a maximum level of impurities (mostly nitrogen) of 10 volume parts per million. A first low-background data taking period with \Xe{136}-enriched gas (hereafter Run-V) was conducted from February 2019 to June 2020, achieving an exposure of 271.6 days. During this run, two gas recoveries took place in order to carry out minor interventions not impacting the detector performance. A second low-background period with \Xe{136}-depleted gas (hereafter Run-VI) was carried out from October 2020 to June 2021, reaching a total run time of 208.9 days. The integrated data acquisition (DAQ) live-times during Run-V and Run-VI are 97.04$\pm$0.01\% and 97.86$\pm$0.01\%, respectively. The trigger efficiency reaches a plateau of 97.6$\pm$0.2\% for events above $\approx$400 keV. The same operation conditions were chosen for the two runs, with gas pressure, drift field and EL field set to $\approx$10.2~bar, 0.4~kV/cm, and 1.3~kV/(cm bar), respectively.  The time evolution of the gas density was monitored, with the largest sources of variability being the re-filling of the detector between the different data taking periods. The integrated electron number density in the gas during Run-VI is 1.9$\pm$0.2\% larger than that in Run-V, inducing a relative increase in the observed event rates of 2.4$\pm$0.6\% (according to MC studies) due to the reduction in the gamma-ray attenuation length and the larger probability of multi-Compton interactions. An uncertainty of 0.2\% in the total number of Xe atoms in the active volume is derived from a 0.5~K uncertainty in the average gas temperature inside the active volume, in turn inferred from the temperature spread among sensors mounted at various locations in the NEXT-White detector and surroundings.

Continuous detector calibration and monitoring were carried out with a \Kr{83\mathrm{m}} low-energy (41.5~keV) calibration source \cite{Martinez-Lema:2018ibw}. The high rate of krypton events induces a typical DAQ dead-time of 2--6\%, which is measured on a daily basis. The electron drift velocity was stable within 1\%, with a value around 0.92~mm/$\mu$s during both runs. The electron drift lifetime ranged from $\approx$5~ms to $\approx$14~ms ($\approx$7~ms to $\approx$14~ms) during Run-V (Run-VI), continuously improving due to the gas recirculation through a MonoTorr PS4-MT50-R SAES heated getter. The electron lifetime values achieved are significantly larger than the maximum drift time of $\approx$0.6~ms, and demonstrate the excellent gas purity conditions achieved with both \Xe{136}-enriched and \Xe{136}-depleted gas. With a light yield of $\approx$300 photo-electrons per keV, the energy resolution at 41.5~keV remained stable around 4\% full width at half maximum (FWHM).

\section{Event reconstruction and selection}

\begin{figure*}
\includegraphics[width=2.0\columnwidth]{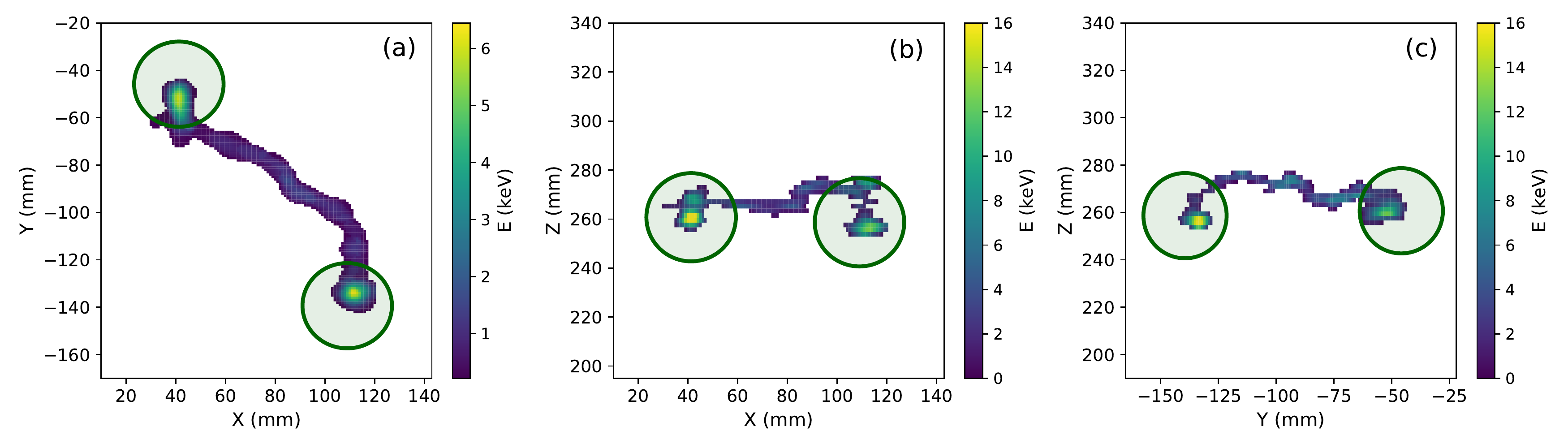}
\includegraphics[width=2.0\columnwidth]{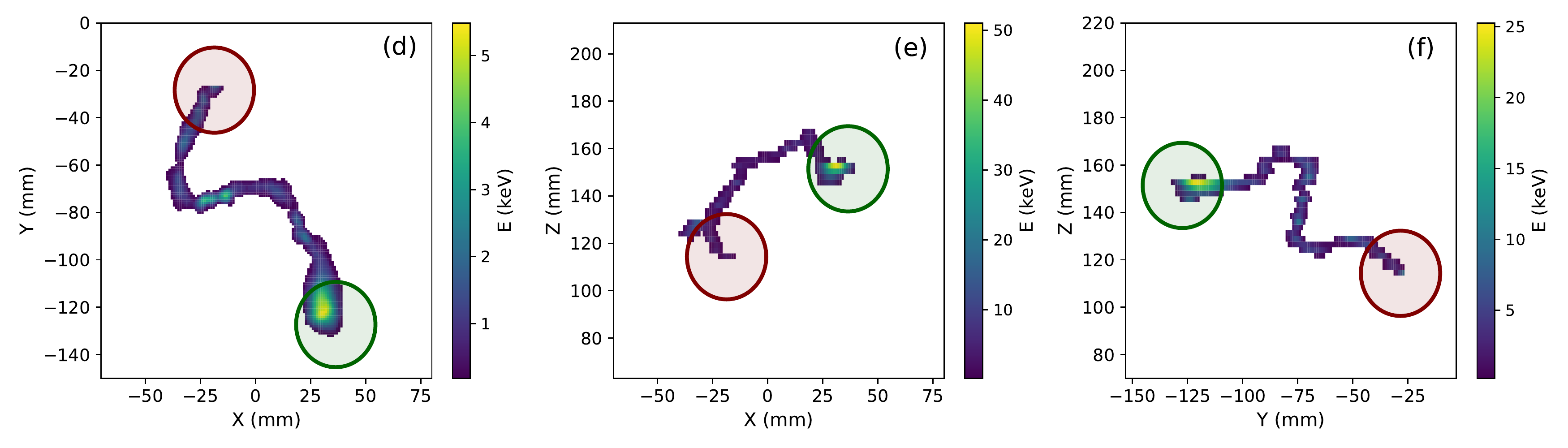}
\caption{\label{fig:display} XY, XZ \& YZ projections of 3D double-electron [panels (a), (b), and (c), respectively] and single-electron [panels (d), (e), and (f)] tracks of 1.7~MeV ($E_\mathrm{b,min}\approx 375$ keV), according to the selection described in the text. The circles mark the energy integration region used to define the track blobs. While two clear Bragg peaks are present in the \bb candidate (blobs of 529 and 446~keV), only one is observed in the single electron track (blobs of 104 and 755~keV).}
\end{figure*}

In the individual reconstruction of triggered events, a first stage detects S1 and S2 signals within the PMT waveforms. The SiPM hits providing the X and Y coordinates are reconstructed separately for each 2$\mu$s slice of the S2 signals. The S2 slice times are converted into Z positions by considering the time difference with respect to the S1 signal in the event. The energy obtained with the PMTs in the same time slice is divided among the reconstructed 3D hits, proportionally to the charge collected by the corresponding SiPMs. The resulting hit energy is corrected by the electron drift lifetime, geometrical effects, and time variations according to \Kr{83\mathrm{m}} data collected within a $\approx$24 h period. A second reconstruction stage is performed in order to reverse the blurring induced by the electron diffusion and the EL light production. A Richardson-Lucy deconvolution is applied to the 3D hits relying on a point spread function derived from \Kr{83\mathrm{m}} events \cite{NEXT:2020try}. The deconvolved 3D hits are then grouped into volume elements of (5~mm)$^3$, which are used to build tracks following the connectivity criteria established by a breadth-first search algorithm \cite{Cormen2001_intro_algorithms}. The energies of the end-points of each track, hereafter `blobs', are defined by integrating the energy of the hits contained within spheres of 18~mm in radius centered in the identified extremes \cite{NEXT:2020try}. Figure~\ref{fig:display} shows examples of two observed tracks of 1.7~MeV.

The event energy $E_\mathrm{evt}$ is estimated by summing the energy of all calibrated hits. The energy scale is calibrated by means of data from \Cs{137} and \Th{228} sources deployed in dedicated ports on the NEXT-White pressure vessel. An empirical second-degree polynomial energy scale model has been adopted, yielding residuals on the peak positions that appear in low-background data (\Co{60}, \K{40}, and \Tl{208}) from 1173 keV to  2615 keV of below 0.3\%. A stable energy resolution of $\approx$1\% FWHM at 2615~keV is found in all calibration campaigns \cite{Renner:2019pfe}.


A two-stage selection procedure is applied to the reconstructed events. First, a fiducial selection to reject backgrounds from detector surfaces and/or with multi-track topologies is performed. We require single-track events (expected for \bb events) to be fully contained within the volume defined by 20$<$Z$<$510~mm and ${\mathrm R}=\sqrt{{\mathrm X}^2+{\mathrm Y}^2}<$195~mm. According to the average gas density, the remaining fiducial mass is 3.50$\pm$0.01~kg. Second, the \bb selection adds two requirements to the fiducial ones: the tracks are required not to have any common hits in their blobs ({\it i.e.}, overlapping blobs), and their less energetic blobs are required  to have a blob energy $E_\mathrm{b}$ greater than a given energy threshold $E_\mathrm{b,min}$. This ensures that the track has two Bragg peaks at the extremes, corresponding to the stopping points of the two electrons. Tracks not fulfilling this condition are flagged as single-electron-like. $E_\mathrm{b,min}$ is defined as a function of the energy of the event, optimized by means of MC studies. The ratio of the signal efficiency over the square root of the background acceptance ranges from $\approx$2.3 to $\approx$3.1 for events between 1 and 3~MeV energy, consistent with Ref. \cite{NEXT:2020try}. According to this selection, the top and bottom panels of Fig. \ref{fig:display} correspond to double-electron and single-electron candidate events, respectively. We only consider $E_\mathrm{evt}>1$~MeV events in the current analysis, because for lower-energy (shorter) tracks the topological discrimination worsens considerably.


The data selection efficiencies of the two selection stages are computed by means of \Tl{208} calibration data, independently for Run-V and Run-VI. The efficiency for the fiducial selection is obtained using all events with  $E_\mathrm{evt}>1$~MeV. The efficiency of the \bb selection is obtained as the product of the fiducial selection efficiency and the no-overlap and blob energy cut efficiencies. While for the no-overlap cut, all the events above 1 MeV are considered, the efficiency of the blob energy cut is evaluated separately for double-electron and single-electron events. For the former topology, only calibration events inside the double-escape peak at 1.6~MeV produced by 2.6~MeV \Tl{208} gamma-rays are used. As discussed in Refs. \cite{NEXT:2019gtz,NEXT:2020try}, events inside (outside) this peak are mostly populated by pair-production (Compton scattering) events. The overall \bb selection efficiency for double-electron (single-electron) calibration tracks of  $E_\mathrm{evt}>1$~MeV energy is measured to be 24.7$\pm$0.5\% (2.24$\pm$0.06\%) in Run-V, and 27.5$\pm$0.6\% (2.34$\pm$0.07\%) in Run-VI. The efficiencies of the single-electron background, the double-electron background, and the \bbtwonu MC samples are adjusted according to the ratios between these measurements and the calibration MC expectations. The energy dependence of the selection cuts observed in data is found to be consistent with the MC expectation.

\section{Radiogenic background}

\begin{figure*}
\includegraphics[width=2.0\columnwidth]{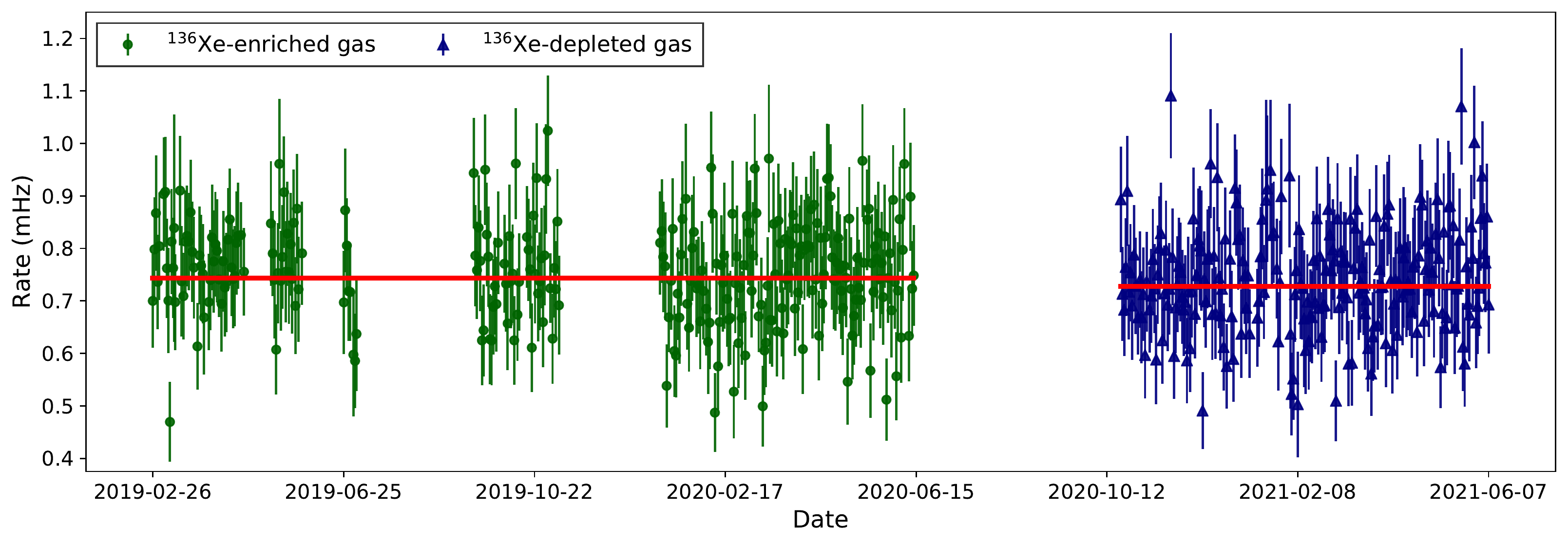} 
\caption{\label{fig:ratevstime}Fiducial event rate along the low-background data taking periods. Green dots (blue triangles) show the daily rates corresponding to the \Xe{136}-enriched (\Xe{136}-depleted) campaign. The horizontal red lines present fits to the data with 0-degree polynomials, yielding $\it{p}$-values of 24\% and 15\% for Run-V and Run-VI, respectively.}
\end{figure*}

The time stability of the backgrounds has been assessed by different means. As shown in Fig. \ref{fig:ratevstime}, the rate evolution of fiducial events is consistent with a constant distribution within Run-V and Run-VI, corresponding to integrated rates of 0.758$\pm$0.006 mHz and 0.742$\pm$0.011 mHz, respectively. The observed difference (0.016$\pm$0.013 mHz) is consistent with the \bbtwonu rate expectation in Run-V ($\approx$0.027 mHz) based on the half-life reported in Ref. \cite{Albert:2013gpz}. To assess the stability of the different background sources, the fiducial events have also been fitted to a radiogenic background model built upon the radiopurity screening of the detector materials, as done in Ref. \cite{Novella:2019cne}. The model consists of the contributions of \K{40}, \Co{60}, \Tl{208}, and \Bi{214}, from 23 different detector volumes. The fit considers both the energy spectrum and the Z distribution of the events, measuring the rate contribution of each isotope from three effective volumes: the cathode, the anode, and any other region. The small contribution of the \Xe{136} \bbtwonu is fixed to the expectation from Ref. \cite{Albert:2013gpz}, and the initial kinematics of the events simulated with the DECAY4 Monte Carlo generator \cite{Ponkratenko:2000um}. The 12 best-fit background contributions are found to be fully consistent between Run-V and Run-VI. Finally, the intensity of the \Co{60} 1173 keV gamma line has been monitored over time. Because no significant variations have been observed, the background induced by this cosmogenic isotope is assumed to be stable. While the radio-impurities in the detector materials are expected to be constant in time, these results discard also the hypothesis of significant time-evolving background sources from the gas system or sizable contributions from \Xe{137} activations.

\section{Measurement of the \bbtwonu half-life}

The measurement of the \Xe{136} \bbtwonu half-life relies on the combination of the Run-V and Run-VI data samples, with Run-VI data providing a measurement of the backgrounds. For our main result, the half-life is derived from a direct background subtraction. The normalization systematic uncertainties account for both the rate subtraction error (considering the DAQ live-time, the gas density, and the selection efficiencies of single and double-electron events in both periods), and the signal normalization error (considering the isotopic composition of the gas, the number of xenon atoms and the trigger efficiency). Within the total normalization uncertainties presented in Table \ref{tab:sys}, the one associated with the selection efficiency of double-electron (2e$^{-}$) and single-electron (1e$^{-}$) tracks ($\approx$2\% and $\approx$3\%, respectively) dominates. Since these efficiencies are derived from independent calibration data samples, they are conservatively assumed to be fully uncorrelated between Run-V and Run-VI. Although an energy-independent energy scale relative uncertainty of 0.3\% has been adopted, as inferred from the residuals obtained from our energy scale model, no significant impact on the results has been observed.

\begin{table}[b]
\caption{\label{tab:sys} Rate normalization uncertainties in Run-V and Run-VI. The last column indicates whether the uncertainty is correlated between the two periods. Sources above the continuous line affect the background-subtracted rate, while the sources below have an impact on the \bbtwonu signal.}
\begin{ruledtabular}
\begin{tabular}{cccccccc}
Source  & Run-V (\%) & Run-VI (\%) & Correlated\\
\hline
DAQ live-time & 0.01 & 0.01 & No \\
Gas density & -\footnotemark[1] & 0.6 & No \\
$\beta\beta$ selection for 2e$^{-}$ & 2.1 & 2.1 & No \\
$\beta\beta$ selection for 1e$^{-}$ & 2.8 & 3.0 & No \\ \hline
\Xe{136}-fraction & 0.4 & 0.2 & No \\
Number of Xe atoms & 0.2 & 0.2 & Yes \\
Trigger efficiency & 0.2 & 0.2 & Yes \\
\end{tabular}
\end{ruledtabular}
\footnotetext[1]{Run-VI corrected with respect to Run-V.}
\end{table}

\begin{figure}
\includegraphics[width=0.99\columnwidth]{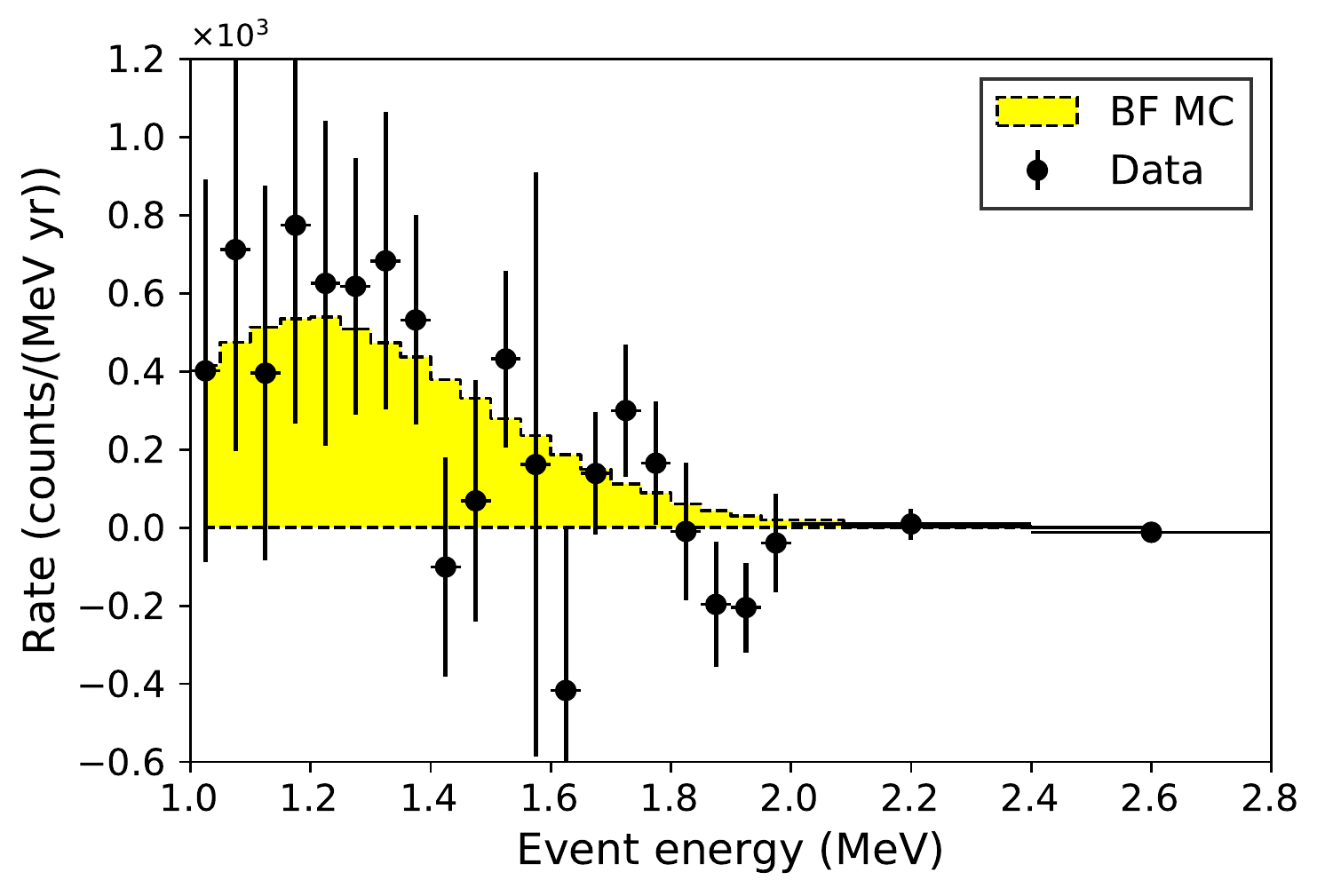}
\caption{\label{fig:bgsubfits} Background-subtraction \bbtwonu fit. The background-subtracted data (black dots) are superimposed to the best-fit MC (yellow histogram). The error bars correspond to the statistical errors in Run-V and Run-VI.}
\end{figure}

Once corrected for the differences in DAQ live-time, gas density and selection efficiencies, the subtraction of the double-electron-like rate in Run-VI to the one observed in Run-V yields R(\Xe{136})=251$\pm$83(stat)$\pm$29(sys) yr$^{-1}$. Thus, a positive \bbtwonu signal is observed at 2.9$\sigma$ from this rate-only measurement. In order to derive the half-life of the \Xe{136} \bbtwonu decay from the background subtracted energy spectrum, a fit is performed to the corresponding MC expectation. In this case, the small \bbtwonu contribution in the Run-VI data is taken into account. The subtraction systematic uncertainty is introduced in the fit as a covariance matrix. The signal normalization uncertainty is decomposed into the uncorrelated (isotopic composition) and correlated (number of xenon atoms and trigger efficiency) contributions between Run-V and Run-VI. Being energy-independent, these errors are introduced in the fit as three nuisance parameters with Gaussian priors. With a $\chi^{2}/\mathrm{dof}$ of 16.1/21 ($\it{p}$-value=76\%), the fit yields a best-fit value for the rate of \bbtwonu events of R(\Xe{136})=291$\pm$73(stat)$\pm$28(sys) yr$^{-1}$. The best-fit rate corresponds to a \bbtwonu half-life of $\Ttwonu=2.34^{+0.80}_{-0.46}\textrm{(stat)}^{+0.30}_{-0.17}\textrm{(sys)}\times10^{21}~\textrm{yr}$. The rejection of the null hypothesis reaches 3.8$\sigma$, while the expected median sensitivity is 4.1$\sigma$ according to the half-life reported in Ref. \cite{Albert:2013gpz}. The background-subtracted \bbtwonu event energy spectrum is presented in Fig. \ref{fig:bgsubfits}. This result is compatible with the two previous measurements in Ref. \cite{Albert:2013gpz} ($\Ttwonu=2.165\pm0.0016\textrm{(stat)}\pm0.059\textrm{(sys)}\times10^{21}~\textrm{yr}$) and Ref. \cite{KamLAND-Zen:2019imh} ($\Ttwonu=2.23\pm0.03\textrm{(stat)}\pm0.07\textrm{(sys)}\times10^{21}~\textrm{yr}$). In an alternative analysis, a consistent $\Ttwonu$ value is also obtained by considering the background-subtracted blob energy distribution instead of the event energy, as summarized in the Appendix \ref{app:blobfit}.

\begin{figure}
\includegraphics[width=\columnwidth]{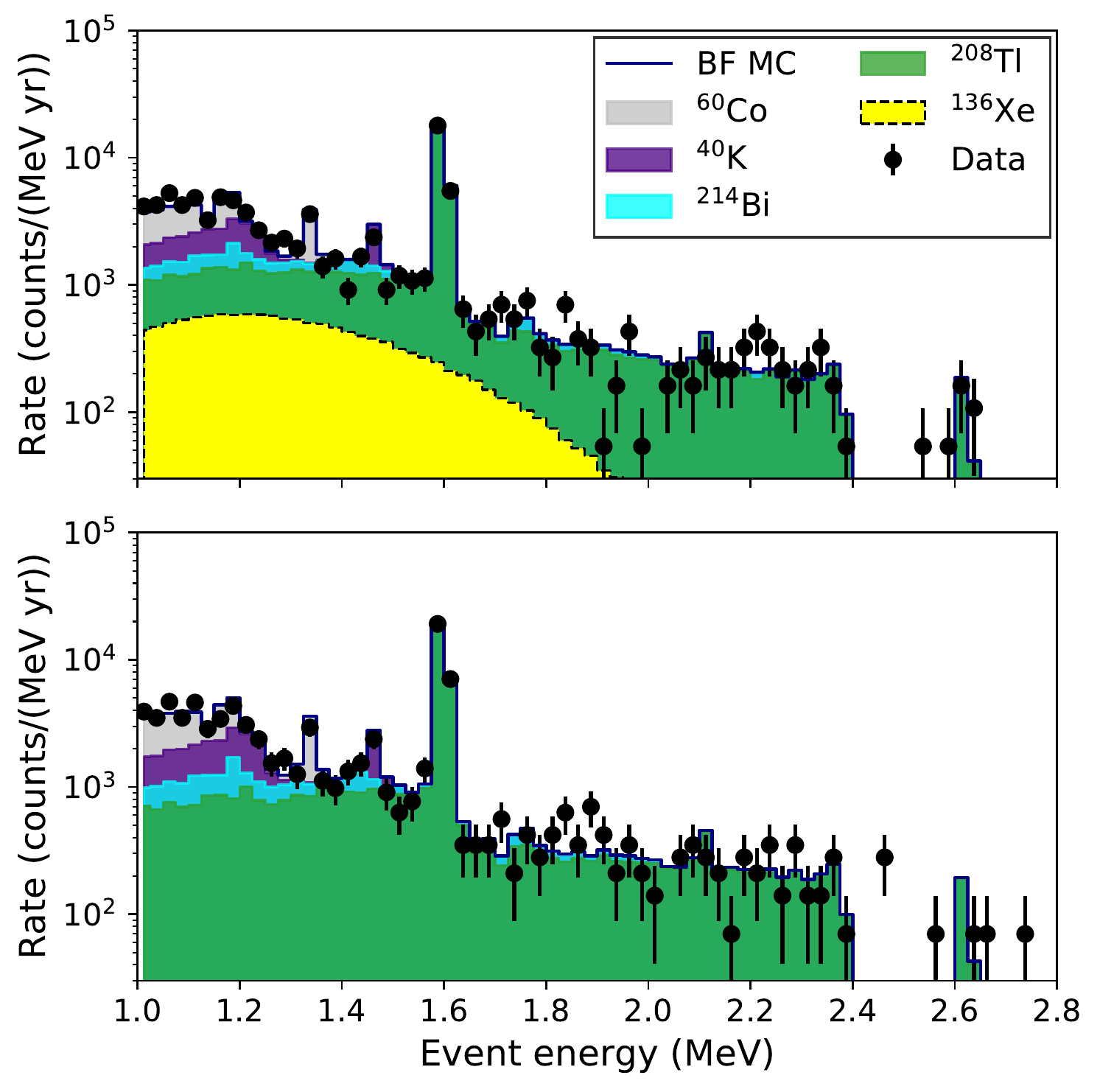}  
\caption{\label{fig:bgmodelfit}Background-model-dependent \bbtwonu fit. \bb-like event rates in Run-V (top) and Run-VI (bottom) are superimposed to the best-fit MC, accounting for \K{40}, \Co{60}, \Tl{208} and \Bi{214} background contributions.}
\end{figure}

A background-model-dependent fit of the event energy has been performed in order to validate the background-subtraction result. In this fit, the \bb-candidates selected in Run-V and Run-VI are jointly fitted to the radiogenic background model. Apart from the rate of \bbtwonu events, the contributions from \K{40}, \Co{60}, \Tl{208}, and \Bi{214} background events are also extracted. The data superimposed to the best-fit MC are shown in Fig. \ref{fig:bgmodelfit}. The best-fit background rates are R(\K{40})=10$\pm$2 $\mu$Hz, R(\Co{60})=14$\pm$2 $\mu$Hz, R(\Tl{208})=40$\pm$2 $\mu$Hz, and R(\Bi{214})=6$\pm$3 $\mu$Hz. The \bbtwonu best-fit rate is R(\Xe{136})=334$\pm$78(stat)$\pm$54(sys) yr$^{-1}$, corresponding to a half-life of $\Ttwonu= 2.14^{+0.65}_{-0.38}\textrm{(stat)}^{+0.46}_{-0.26}\textrm{(sys)} \times10^{21}~\textrm{yr}$ (4.1$\sigma$ significance). The goodness of fit, $\chi^2/\mathrm{dof}$ = 146.1/114 ($\it{p}$-value = 2.3\%), reveals some limitations in the simulation. However, the small difference in the best-fit \Ttwonu with respect to the background-subtraction fit indicates that no significant bias is induced.

\section{Conclusions}

In summary, the operation of the NEXT-White detector with \Xe{136}-enriched and \Xe{136}-depleted xenon gas has enabled the measurement of the \bbtwonu half-life of \Xe{136}, using a fiducial mass of only $\approx$3.5 kg. The analysis relies on two unique capabilities of the NEXT technology, namely, the topological signature of the events and the direct subtraction of backgrounds. This background subtraction technique, novel in the field, offers results with very small dependence on the Monte Carlo assumptions. A similar approach may be exploited to conduct background-model-independent \bbnonu searches in current- and future-generation detectors, such as xenon time projection chambers or loaded liquid scintillator detectors.

\acknowledgments
The NEXT Collaboration acknowledges support from the following agencies and institutions: the European Research Council (ERC) under Grant Agreement No.\ 951281-BOLD; the European Union's Framework Programme for Research and Innovation Horizon 2020 (2014--2020) under Grant Agreement No.\ 957202-HIDDEN; the MCIN/AEI/10.13039/501100011033 of Spain and ERDF A way of making Europe under grant RTI2018-095979, the Severo Ochoa Program grant CEX2018-000867-S and the Mar\'ia de Maeztu Program grant MDM-2016-0692; the Generalitat Valenciana of Spain under grants PROMETEO/2021/087 and CIDEGENT/2019/049; the Portuguese FCT under project UID/FIS/04559/2020 to fund the activities of LIBPhys-UC; the Pazy Foundation (Israel) under grants 877040 and 877041; the US Department of Energy under contracts number DE-AC02-06CH11357 (Argonne National Laboratory), DE-AC02-07CH11359 (Fermi National Accelerator Laboratory), DE-FG02-13ER42020 (Texas A\&M), DE-SC0019054 (Texas Arlington) and DE-SC0019223 (Texas Arlington); the US National Science Foundation under award number NSF CHE 2004111; the Robert A Welch Foundation under award number Y-2031-20200401. DGD acknowledges support from the Ram\'on y Cajal program (Spain) under contract number RYC-2015-18820. Finally, we are grateful to the Laboratorio Subterr\'aneo de Canfranc for hosting and supporting the NEXT experiment.

\appendix

\section{Blob energy fit}
\label{app:blobfit}

This appendix describes the methods and results of the alternative \bbtwonu analysis where the background-subtracted distribution of the energy of the less energetic blob in the track (blob energy, in the following) is fitted instead of the event energy. The event reconstruction is the same as for our main analysis, while the selection of $E_\mathrm{evt}>1$~MeV events differs in two ways. First, the blob energy cut $E_\mathrm{b}>E_\mathrm{b,min}$ is {\it not} applied. This provides a larger statistical sample with respect to the \bb selection, but less signal-enriched. Second, events in the \Tl{208} double escape peak (1.550 $<$ $E_\mathrm{evt}$ $<$ 1.615 MeV) are rejected, in order to suppress the irreducible double-electron background from gamma-ray pair production interactions. Prior to their subtraction and fitting, \Xe{136}-enriched (Run-V) and \Xe{136}-depleted (Run-VI) rates are corrected for differences in DAQ live time, gas density and selection efficiencies. The first two corrections (DAQ live time and gas density) are identical to the ones applied to our main analysis, with uncertainties listed in Table~\ref{tab:sys}. Because of the two above-mentioned differences in event selection, the associated corrections are also different, with 0.3\% (0.4\%) uncertainties for Run-V (Run-VI), uncorrelated between the two runs. Overall, the rate normalization systematic uncertainty affecting the background-subtracted rate is 0.9\%. A calibration procedure is also applied to equalize the blob energy scale for Run-V, Run-VI and MC simulated events, separately for single-electron and double-electron events, using \Tl{208} calibration data. Four uncorrelated blob energy scale systematic uncertainties are assigned, for Run-V single-electron (0.5\%), Run-V double-electron (2.1\%), Run-VI single-electron (0.4\%), and Run-VI double-electron (2.1\%) events, respectively.

\begin{figure}
\includegraphics[width=0.99\columnwidth]{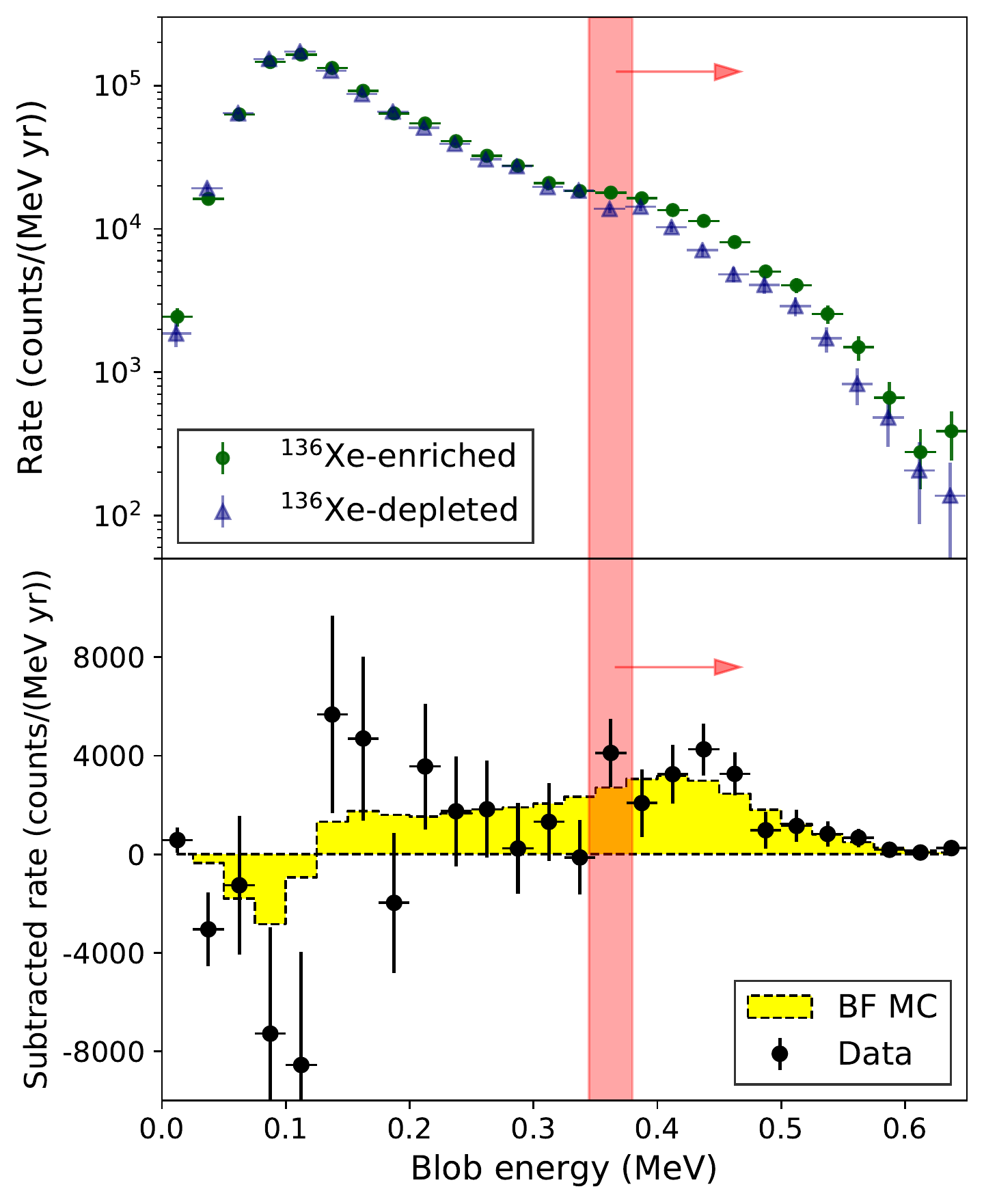}
\caption{\label{fig:eblob_fit} Top: Rates as a function of blob energy for the \Xe{136}-enriched and \Xe{136}-depleted datasets. Bottom: The background-subtracted rate versus blob energy is shown with statistics-only error bars, together with the best-fit MC prediction (yellow histogram). The red band shows the energy-dependent threshold applied in the \bb selection considered in the event energy fit.}
\end{figure}

The top panel in Fig.~\ref{fig:eblob_fit} compares the Run-V and Run-VI rates as a function of blob energy, after applying the small corrections and calibrations mentioned above. In both datasets, the rates are dominated by single-electron background events with $E_\mathrm{b}\approx$100~keV. The secondary bumps at 300--550 keV are due to double-electron background events (Run-V and Run-VI) and to the \bbtwonu signal (Run-V only). The bottom panel in Fig.~\ref{fig:eblob_fit} shows the background-subtracted (Run-V minus Run-VI) rate, superimposed with the best-fit MC prediction. Together with the \bbtwonu rate parameter, the fit incorporates five additional nuisance parameters affecting the MC predictions. The nuisance parameters account for the rate normalization systematic uncertainty and for the four rate shape systematic uncertainties described above. With a $\chi^2/\mathrm{dof}$=24.8/25 ($\it{p}$-value of 47\%), the fit yields a rate of \bbtwonu events of R(\Xe{136})=825$\pm$122(stat)$\pm$94(sys) year$^{-1}$ at 68\% confidence level. The significance of a non-zero \Xe{136} rate measurement is 5.4$\sigma$, to be compared with a 4.2$\sigma$ expected significance assuming the half-life value reported in Ref.  \cite{Albert:2013gpz}.

From the fitted \bbtwonu rate, the measured \Xe{136} isotopic fractions, the average number of Xe atoms in the active volume during Run-V ($(1.909\pm 0.004)\times 10^{25}$) and the overall efficiency to select a \bbtwonu decay in the active volume ($(11.72\pm 0.02)\%$), we obtain a measured half-life of $\Ttwonu=1.66^{+0.29}_{-0.21}\textrm{(stat)}^{+0.25}_{-0.15}\textrm{(sys)}\times10^{21}~\textrm{year}$. This measurement is in agreement with our main result based on the \bb selection and event energy fitting.

\bibliography{apssamp}

\end{document}